\documentclass[rnote, printer]{aa}

\usepackage{amsmath}
\usepackage{txfonts}
\usepackage{natbib}
\usepackage{graphicx}

\newcommand{\bdes}{\begin{description}}
\newcommand{\edes}{\end{description}}

\begin{document}

\title{On the pressure of collisionless particle fluids}
\subtitle{The case of solids settling in disks}

\author{F. Hersant\inst{1,2} }

\offprints{Franck Hersant \email{Franck.Hersant@obs.u-bordeaux1.fr}}

\institute{
Universit\'e de Bordeaux, Laboratoire d'Astrophysique de Bordeaux (LAB)
\and
CNRS/INSU - UMR5804 ; BP 89, F-33271 Floirac Cedex, France
}

\date{Received ??? / Accepted ???}

\abstract
{}
{Collections of dust, grains, and planetesimals are often treated as a pressureless fluid. We study the
validity of neglecting the pressure of such a fluid by computing it exactly for the case of
particles settling in a disk.}
{We solve a modified collisionless Boltzmann equation for the particles and compute the corresponding moments
of the phase space distribution: density, momentum, and pressure.}
{We find that whenever the Stokes number, defined as the ratio of the gas drag timescale to the orbital
timescale, is more than $1/2$, the particle fluid cannot be considered as pressureless. While we show it
only in the simple case of particles settling in a laminar disk, this property is likely to remain true for most flows,
including turbulent flows.}
{}


\maketitle

\section{Introduction}

The interaction of solid particles with gas in protoplanetary disks is one of the key mechanisms for
planetesimals and planet formation. Moreover, this fundamental process gets more and more attention for other
astrophysical media (interstellar medium, planetary atmospheres, etc.) as grains can dampen fluid
motions and be the targets of adsorption and condensation of molecules and can be the base for complex
chemistry, e.g., the formation of the molecule H$_2$ in the interstellar medium. With the advance of
observational facilities and techniques, dust dynamics is understood more and more as a crucial process for
most media, while, due to the discrete nature of grains, modelling their dynamics remains tricky.  \\

There are two main methods for computing the dynamics of particles embedded in a fluid.
One can either study a large number of individual
particles using N-body techniques or consider the particles as a fluid, which is treated with a
modified Navier-Stokes/Euler equation. While the first method is reliable and precise, 
it can only be applied to a finite (and in fact limited) number of particles; therefore, macroscopic
quantities reconstructed from the N-Body simulations outputs are sometimes inaccurate due to a low sampling of
the phase space. 
The
second method enables one to consider an almost infinite ensemble of particles by computing their
macroscopic properties (density, velocity field, etc.). This makes the ``two-fluid" approach
(gas and solids) the preferred method of treating the dynamics and growth of solid particles in a fluid,
especially in the planet formation community. Consequently, 
a dense and rich literature is based on this approach, and many fundamental mechanisms of planetary formation have been
unraveled \citep[e.g.][]{Cuzzi93,Dubrulle95,Stepinski96,Barriere05,Inaba05,Fromang06,Johansen06}.\\

The two-fluid approach is usually implemented by introducing two Euler equations, coupled through the gas drag
force. Moreover, in the fluid equation for particles, the pressure term is usually ignored.
This approximation is often justified by the statement that collisions between particles and gas are much more
probable than collisions between particles making the pressure term negligible compared to
the gas drag force. \\

The purpose of the present paper is to question the validity of ignoring the pressure of the particle fluid by
computing it in the very simple case of particles settling in a disk. The vertical dynamics of a solid body in
a disk can be understood as a damped oscillator. In a Keplerian disk, if gas is absent or very tenuous, the particle oscillates
around the midplane at a frequency equal to the orbital frequency (the oscillation comes from inclined orbits).
The friction with the gas damps this oscillation, as the particle feels a ``headwind" when it crosses the
gas disk. Like any damped oscillator, there are two regimes: an
underdamped regime for low damping and an overdamped regime, where the particle has no time to oscillate for a full
period. This occurs for high damping rates. \\

We can compute the pressure of the solid fluid embedded in a laminar gas disk by solving
exactly the collisionless Boltzmann (or Vlasov) equation for the phase space density of particles. As we 
show, as soon as the particles are larger than some critical size, the pressure cannot be neglected. 
This work extends the work of \cite{Garaud04}, who comes to similar conclusions based on asymptotic calculations.\\

In Sect. 2 is developed the general formalism. Section 3 is devoted to a very simple test case that can be
used to check the validity of more general solutions and intuit how the pressure varies. In Sect. 4, 
the exact general solution of the Vlasov equation is given. In Sect. 5, this solution is applied to a special
case that retains most of the expected general properties. A surprisingly simple solution is found for the
pressure and the scaling laws, and asymptotics are discussed. In Sect. 6, we discuss our results and their
implications for more complex and turbulent flows and give our conclusions.

\section{General formalism}

We consider the equation for the evolution of the phase space density in 1+1 dimensions (one in space, one
in velocity) under the action of a drag force from an inert fluid. This is not exactly the common
collisionless Boltzmann or Vlasov equation, as the drag force is not conservative. Instead, 
the general Vlasov equation with non-conservative forces takes the form \citep[see
e.g.][]{Lutsko97,Carballido06, Youdin07}:
\begin{equation}
\frac{\partial f}{\partial t}+u\frac{\partial f}{\partial z} + \frac{\partial}{\partial
u}\left(\frac{F}{m}f\right)=0
\label{vlasov}
\end{equation}
where $z$ is the space coordinate, $u$ is the velocity coordinate, $F/m$ the non-conservative force per
mass (acceleration), and $f$ the phase space density (or probability density function). When $f$ is
known, the macroscopic quantities (density, velocity, energy) are moments of the distribution $f$. In
particular, the
zeroth order moment is the density $n(z,t)$:
\begin{equation}
n(z,t)=\int_{-\infty}^{+\infty} f(z,u,t) du, 
\label{ndef}
\end{equation}
the first-order moment is the momentum (the product of the density by the macroscopic velocity $\tilde{u}$)

\begin{equation}
n(z,t)\ \tilde{u}(z,t)=\int_{-\infty}^{+\infty} u f(z,u,t) du, 
\label{udef}
\end{equation}
and the second-order moment is related to the pressure (a scalar in 1D):

\begin{align}
\nonumber
P(z,t)&=\int_{-\infty}^{+\infty} (u-\tilde{u})^2 f(z,u,t) du\\
&=\int_{-\infty}^{+\infty} u^2 f(z,u,t)
du - n\tilde{u}^2.
\label{Pdef}
\end{align}

In the case of particles settling in a disk, there are two forces: gravity and gas drag. When the
disk is geometrically thin, the vertical gravity can be written as $\frac{F}{m}=-\omega^2 z$, where $\omega$ is the orbital
frequency, while the drag force, in the absence of gas motions, can be written as $\frac{F}{m}=-u/\tau$ where
$\tau$ is the so-called stopping time \citep{Whipple72}. This is the characteristic timescale for a particle to stop
in a fluid. Here, we consider $\tau$ as a free parameter, which would in general depend upon
the particle size, morphology, density, the density of the gas, and the turbulent intensity in the grain
trail. Here, we consider it as independent of both space and velocity. This is an important
simplification, corresponding to the so-called Epstein regime, where grains are smaller than the mean free
path of the underlying gas. It should, however, be noted that the mean free path $l_{mfp}$ in a circumstellar disk is very
large, typically varying like $l_{mfp}=10^8 \left(\frac{R}{100\ \mathrm{AU}}\right)^{5/2}$ cm, assuming that
the gas number density $n_g$ varies like $n_g=10^8 \left(\frac{R}{100\ \mathrm{AU}}\right)^{-5/2}$ cm$^{-3}$
\citep{Guilloteau98,Hersant05}.\\

Under these conditions, Eq.
(\ref{vlasov})
becomes

\begin{equation}
\partial_t f +u\partial_z f - \partial_u \left( \omega^2 z +\frac{u}{\tau} \right) f=0
\end{equation}
where $z$ is now the vertical space variable and $u$ the vertical velocity. To simplify the notations, we use the
convention $\partial_x=\frac{\partial}{\partial x}$. This equation can be developed
into pure advective terms:

\begin{equation}
\partial_t f +u\partial_z f - \left( \omega^2 z +\frac{u}{\tau} \right) \partial_u f -\frac{f}{\tau}=0
\end{equation}
The last term on the lefthand side ($-f/\tau$) is the difference with a Vlasov equation for conservative forces and deserves
some comments. This term induces an exponential growth of $f$. Without solving the equation, one can 
understand that the conservation of the number of particles (following from Eq. \ref{vlasov}),
requires that the surface of the phase space field of density $f$ has to collapse in one direction so
that, at infinite times, the field of $f$ tends to a discontinuous distribution, consisting of a family of
curves, but nothing like a continuous surface. This property of the equation remains valid even in 3+3 dimensions and with gas
turbulence.\\

At this point, it is convenient to adimension the time, introducing two new variables $t'=\omega t$ and
$u'=u/\omega$. Using these new variables and omitting the primes for clarity, we get

\begin{equation}
\partial_t f +u\partial_z f - \left(z +\frac{u}{S} \right) \partial_u f -\frac{f}{S}=0
\end{equation}
where $S=\omega \tau$ is a Stokes number of the problem. For small (large) particles, $S$ is
small (large) due to the strong (weak) coupling of the particles with the fluid.\\

We now can artificially get rid of the exponential growth term by defining

\begin{equation}
\psi=f\ e^{-t/S}
\end{equation}
and finally get

\begin{equation}
\partial_t \psi +u\partial_z \psi - \left( z +\frac{u}{S} \right)\partial_u \psi=0
\label{psi}
\end{equation}
The behavior of this equation is rather easy to foresee. For infinite $S$, the remaining terms correspond
to a rotation in phase space. This rotation is a direct consequence of energy conservation. In real
space, particles oscillate around the midplane with zero velocity at the highest absolute altitude and
maximum velocity when particles cross the midplane. The gas-drag term induces a collapse along the velocity
axis. The net outcome of both the rotation and the collapse along the velocity axis (which corresponds in
the real space to the dampening by gas drag of the inclination of the particle orbit) is a global collapse
towards the center of the phase space ($z=0$, $u=0$), but not a spherical collapse.

\section{Simple case}

Although the general solution is given in the next section, it is useful to solve Eq.
(\ref{psi}) in the simplest case: infinitely large or massive particles ($S=+\infty$). In this case, the
equation admits a simple stationary solution with separated variables:  
\begin{equation}
\psi=Ke^{-\frac{z^2+u^2}{2\sigma^2}}.
\label{psieasy}
\end{equation}
Using the definition of the moments, the pressure is straightforward:

\begin{equation}
P(z)=n(z) \sigma^2.
\label{Peasy}
\end{equation}
where the density is
\begin{equation}
n(z)=K\sigma \sqrt{2\pi}e^{-\frac{z^2}{2\sigma^2}}.
\label{neasy}
\end{equation}

Equation (\ref{Peasy}) is similar to an equation of state for isothermal fluids, where $\sigma$ plays the
role of a (constant) sound speed. As we will see in Sect. \ref{2gauss}, this behavior remains valid even
in more general cases. With the equation of state (\ref{Peasy}), solution (\ref{neasy}) is the result of the
hydrostatic equilibrium of an isothermal gas, the pressure force balancing the vertical gravity. The stationary
solution (\ref{psieasy})
being even in $u$, it
corresponds to a zero macroscopic velocity. In this case, where particles have an infinite inertia, the
pressure term is thus dominant in the structure of the particle disk.

\section{General solution}

Equation (\ref{psi}) can be solved in the general case, using standard techniques (the method of characteristics for
example). We will not develop this here as it is easier to check that the following solution is the general
solution of Eq. (\ref{psi}):

\begin{equation}
\psi(z,u,t)=\phi\left(z_0(z,u,t),u_0(z,u,t)\right)
\end{equation}
where $\phi$ is the initial condition and ($z_0$, $u_0$) is the initial phase space position of a particle
located in ($z$, $u$) at a time $t$. The initial positions are given by

\begin{equation}
\begin{cases}
z_0=f_1(t) z\ e^{\alpha t} - f_2(t) u\ e^{\alpha t}\\
u_0=f_2(t) z\ e^{\alpha t} + f_3(t) u\ e^{\alpha t},
\end{cases}
\end{equation}
where $f_1$, $f_2$ and $f_3$ are time dependent functions given by

\begin{equation}
\begin{cases}
f_1(t)=\mathrm{ch}(\beta t)-\frac{\alpha}{\beta}\mathrm{sh}(\beta t)\\
f_2(t)=\frac{1}{\beta}\mathrm{sh}(\beta t)\\
f_3(t)=\mathrm{ch}(\beta t)+\frac{\alpha}{\beta}\mathrm{sh}(\beta t)
\end{cases}
\end{equation}
and $\alpha$ and $\beta$ are parameters depending on the Stokes number $S$:

\begin{equation}
\begin{cases}
\alpha=\frac{1}{2S}\\
\beta=\sqrt{(\alpha+1)(\alpha-1)}=\sqrt{\alpha^2-1}.
\end{cases}
\end{equation}

Here, $\alpha$ is always real and positive, while $\beta$ is either a positive real number (for small Stokes
numbers), or a purely imaginary number (for large Stokes numbers). However, whatever $\beta$, the time functions
$f_1$, $f_2$, and $f_3$ are always real.

\section{Particular case for the initial conditions}
\label{2gauss}

Computing the moments of the phase space distribution cannot be done without the initial conditions.
Here, we choose an initial condition composed of Gaussians, with one Gaussian per dimension, which is relevant to most physically meaningful
cases:

\begin{equation}
\phi(z,u)=Ke^{-\frac{z^2}{2\sigma_z^2}}e^{-\frac{u^2}{2\sigma_u^2}}
\label{initphi}
\end{equation}
where $\sigma_z$ is the initial thickness of the particle sub-disk and $\sigma_u$ is the initial
(presumably small) velocity
dispersion. The normalization constant $K$ is related to the surface density $\Sigma$ by $\Sigma=2\pi K \sigma_z
\sigma_u$. This initial condition assumes that the particles initially have a vanishing mean velocity, a
velocity dispersion and are randomly distributed vertically, following a Gaussian distribution in this
case. Other choices would lead to similar results unless the symmetry with respect to the disk midplane is
broken (particles initially located only in the upper half of the disk, for example).

\subsection{The corresponding moments}

Using definition (\ref{ndef}), the density corresponding to the initial conditions (\ref{initphi}) can be
written as

\begin{equation}
n(z,t)=Ke^{2\alpha
t}\mathrm{exp}\left(-\frac{z^2f_1^2(t)}{2\sigma_z'^2(t)}\right)\mathrm{exp}\left(-\frac{z^2f_2^2(t)}{2\sigma_u'^2(t)}\right)L(z,t)
\end{equation}
where the exponential dependencies can be included into the Gaussian widths:

\begin{equation}
\begin{cases}
\sigma_z'=\sigma_z e^{-\alpha t}\\
\sigma_u'=\sigma_u e^{-\alpha t}
\end{cases}
\end{equation}
and $L(z,t)$ is a function defined as

\begin{equation}
L(z,t)=\int_{-\infty}^{+\infty}
\mathrm{exp}\left(-\frac{u^2f_2^2(t)}{2\sigma_z'^2(t)}\right)\mathrm{exp}\left(-\frac{u^2f_3^2(t)}{2\sigma_u'^2(t)}\right)e^{-pu}du
\end{equation}
with

\begin{equation}
p(z,t)=-\frac{f_1f_2z}{2\sigma_z'^2}+\frac{f_2f_3z}{2\sigma_u'^2}.
\label{p}
\end{equation}

Let us define
\begin{equation}
\lambda=\frac{\sigma_z'\sigma_u'}{\sqrt{\sigma_z'^2f_3^2+\sigma_u'^2f_2^2}}.
\label{lambda}
\end{equation}
Then:

\begin{equation}
L_p=\int_{-\infty}^{+\infty}e^{-\frac{u^2}{2\lambda^2}}e^{-pu}du.
\end{equation}
Here we keep the $p$-dependence explicitly as it will be useful in the following. We have $L(z,t)=L_p$ if $p$
takes the value given by equation (\ref{p}).
This expression can be now written as

\begin{equation}
L_p=F(p)+F(-p)
\end{equation}
where $F(p)$ is the Laplace transform of a Gaussian:
\begin{equation}
F(p)=\int_{0}^{+\infty}e^{-\frac{u^2}{2\lambda^2}}e^{-pu}du=\lambda\sqrt{\frac{\pi}{2}}e^{\frac{\lambda^2p^2}{2}}\mathrm{erfc}\left(\frac{\lambda
p}{\sqrt{2}}\right).
\end{equation}
Therefore, using the {\bf identity} $\mathrm{erfc}(x)+\mathrm{erfc}(-x)=2$, we have
\begin{equation}
L_p=\lambda\sqrt{2\pi}e^{\frac{\lambda^2p^2}{2}}
\label{Lp}
\end{equation}

Using the definition (\ref{udef}) and a similar procedure, we get for the first-order moment:

\begin{equation}
n\tilde{u}=Ke^{2\alpha
t}\mathrm{exp}\left(-\frac{z^2f_1^2(t)}{2\sigma_z'^2(t)}\right)\mathrm{exp}\left(-\frac{z^2f_2^2(t)}{2\sigma_u'^2(t)}\right)H(z,t), 
\end{equation}
where

\begin{equation}
H(z,t)=\int_{-\infty}^{+\infty} u\ e^{-\frac{u^2}{2\lambda^2}}e^{-pu}du.
\end{equation}
Now $H_p$ can be written as a function of $F_p$, using basic properties of the Laplace transform:

\begin{equation}
H_p=\partial_p F(-p)-\partial_p F(p)=-\partial_p L_p.
\end{equation}
Hence,

\begin{equation}
\tilde{u}=-\frac{\partial_p L_p}{L_p}.
\end{equation}

Applying the same procedure to definition (\ref{Pdef}), we get the following expression for the particle fluid
pressure:

\begin{equation}
P=n\frac{\partial_p^2L_p}{L_p}-n\left(\frac{\partial_pL_p}{L_p}\right)^2,
\end{equation}
and using Eq. \ref{Lp}, we finally get the simple expression for the pressure:
\begin{equation}
P(z,t)=n(z,t)\lambda^2(t).
\end{equation}
Here $\lambda$ can be understood as the velocity dispersion of the particles, and its expression
\ref{lambda} generalizes the asymptotics of \cite{Garaud04}.\\

It is interesting to note that the general pressure only depends on $z$ through the density. The Euler
equation for the particle fluid involves only gradients of the pressure so this gradient is in fact directly
proportional to the density gradient. In other words, the particles seem to follow a barotropic equation of
state. This behavior seems closely related to the assumption of constant stopping time \citep{Garaud04}.

\subsection{Scaling and asymptotics}

\subsubsection{Case $\beta$ real (small particles)}

When $\beta$ is real (and positive), $\alpha$ is large, it is easy to check that $\lambda$ is a
decreasing function of time. Its initial value is

\begin{equation}
\lambda(0)=\sigma_u,
\end{equation} 
and the large time asymptotics are
\begin{equation}
\lambda(t \to +\infty) \sim \frac{\beta
\sigma_z}{\sqrt{1+\left(\frac{\sigma_z}{\sigma_u}\right)^2\left(\alpha+\beta\right)^2}}e^{-(\alpha+\beta)t}.
\end{equation}

Starting from a presumably very low value (the initial velocity dispersion of particles), $\lambda$ decreases
with time, asymptotically like an exponential. The location where the pressure decreases the least is the
midplane, since the density tends to a Dirac function when times goes to infinity,

\begin{equation}
n(z=0,t)=K \lambda\sqrt{2\pi}e^{2\alpha t},
\label{middens}
\end{equation}
and

\begin{align}
P(z=0,t)&=\lambda^3 K \sqrt{2\pi} e^{2\alpha t}\\
&\sim\left(\frac{\beta
\sigma_z}{\sqrt{1+\left(\frac{\sigma_z}{\sigma_u}\right)^2\left(\alpha+\beta\right)^2}}\right)^3
e^{-(\alpha+3\beta)t}.
\end{align}

Even in the midplane, the pressure quickly drops with time, starting from a low value. In this case, as has
long been known \citep[see e.g.][]{Cuzzi93,Dubrulle95,Garaud04}, it is safe to neglect the pressure of the
particle fluid. Consequently, particles steadily sediment towards the disk midplane at a velocity close to the
terminal velocity $w^\infty=\omega^2z\tau$ (defined as the asymptotic velocity of any body free-falling in a
fluid), which is the asymptotic velocity of both the particle fluid and individual particles here.
 
\subsubsection{Case $\beta$ imaginary (large particles)}

We define $\gamma$ such that $\beta=i \gamma$. In such a case, $\lambda$ can be written as
\begin{equation}
\lambda=e^{-\alpha t}\frac{\sigma_u}{\sqrt{\left( \mathrm{cos}\ \gamma t+\frac{\alpha}{\gamma}\mathrm{sin}\ \gamma
t\right)^2+\left(\frac{\sigma_u}{\sigma_z}\right)^2\frac{1}{\gamma^2}\mathrm{sin}^2\ \gamma t}}
\end{equation}
For small initial velocity dispersions, the location of the maxima of $\lambda$ corresponds to
\begin{equation}
\mathrm{tan}\ \gamma t=-\frac{\alpha}{\gamma}
\end{equation}
and the amplitude of these maxima is
\begin{equation}
\lambda_{max}(t)=\sigma_z \frac{\gamma}{\alpha}\sqrt{\alpha^2+\gamma^2}\ e^{-\alpha t}.
\end{equation}

It is crucial to note here that the amplitude of the maxima of the time-dependent velocity dispersion $\lambda$ is independent of the initial
velocity dispersion $\sigma_u$. The velocity dispersion is created and maintained by the vertical gravity.  \\

Since the midplane density is still given by Eq. (\ref{middens}), we get the amplitude of the midplane pressure
maxima:

\begin{equation}
P_{max}(t)=K\sqrt{2 \pi}\left(\sigma_z \frac{\gamma}{\alpha}\sqrt{\alpha^2+\gamma^2}.
\right)^3 e^{-\alpha t}
\end{equation}
These pressure maxima have a much lower decreasing rate than in the non-oscillating case, both because it
involves only $\alpha$ here and because $\alpha$ has a lower value.\\

\begin{figure}
\centering
\hspace*{0.2cm}\includegraphics[angle=270,width=0.52\textwidth]{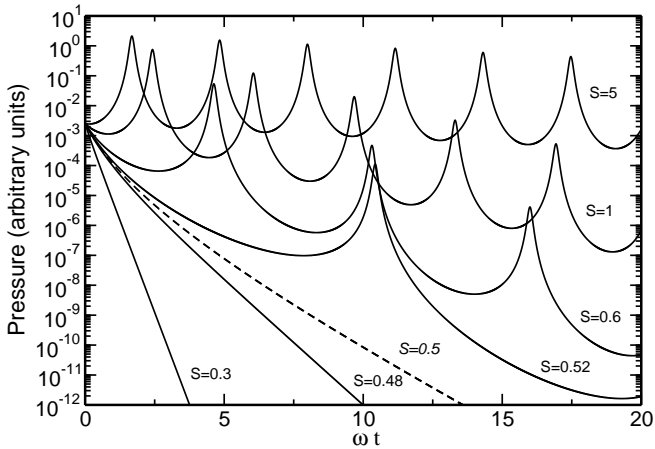}
\caption[]{Pressure of the particle fluid in the midplane as a function of the dimensionless time ($\omega
t$), for different values of the Stokes number $S$
}
\label{pressdisp}
\end{figure}

The midplane pressure as a function of time for different values of the Stokes number $S$ is displayed in
Fig. \ref{pressdisp}. The transition between the case with oscillation $S > 1/2$ and the overdamped case $S <
1/2$ appears very clearly. The pressure is computed for a ratio $\frac{\sigma_u}{\sigma_z}=0.1$. This ratio
defines the pressure at $t=0$ and the width of the maxima, but neither the amplitude of the maxima nor the
decreasing rate.\\

Figure \ref{pressdisp} illustrates that, while the pressure of the solids fluid can be safely neglected for small
values of the Stokes number $S$, the situation is drastically different when the Stokes number reaches
$S=1/2$. When particles oscillate around the midplane, the fluid macroscopic velocity is different from
the terminal velocity $w^\infty$ as the macroscopic pressure force slows down the vertical contraction of the
particle fluid. In other words, the velocity of the particle fluid is not that of individual particles and the fluid
has a nonvanishing internal energy (contrary to the case for small particles).

\section{Discussion and conclusions}

We have shown that, in general, the pressure of the particle fluid cannot be neglected as soon as their Stokes
number is more than $0.5$. The particles seen as a fluid can have an internal energy, and the pressure
force can partly balance gravity. In the extreme case where particles have an infinite inertia, the structure
of the particle fluid is in fact hydrostatic. The existence of pressure is completely independent of the presence or absence
of collisions between solid particles. Instead, the pressure is the result of a velocity dispersion. While
velocity dispersion can be the result of collisions, 
it can also be created dynamically when external forces excite the velocity dispersion (gravity in the
present case) or when the flow itself is converging locally.
Even though our results concern only a very simple and specific problem, they have implications for
a much broader variety of astrophysical and non-astrophysical flows.\\

The Stokes number can be defined in a more general way as the ratio of a dynamical timescale, creating
velocity differences between the solids and the underlying fluid (gas or liquid), and the stopping time, the
characteristic timescale for the particles to reach the fluid velocity. There are many processes able to create
velocity differences between the gas and solids, gravity being only one of them. For example, when two
rivers join, particles embedded in the rivers do not see the confluence like the water does and this can create
a velocity difference between water and the particles, and a velocity dispersion in the particle flow at the
confluence.\\

For particles embedded in a turbulent flow, it is common to define a scale-dependent Stokes number by $S_\eta=
\tau/t_\eta$, where $t_\eta$ is the turnover time of a turbulent eddy of size $\eta$. This Stokes number depends on the
size of the turbulent eddy and scales as $S_\eta \varpropto \eta^{2/3}$ for Kolmogorov-like turbulence
\citep[see e.g.][]{Cuzzi01}. The smallest reachable eddy is dependent on the strength of turbulence,
quantified by the Reynolds number of the flow $Re=\frac{UL}{\nu}$, where $U$ and $L$ are characteristic
velocities and length scale, respectively, and $\nu$ is the molecular kinematic viscosity of the fluid. For
Kolmogorov turbulence, the smallest turbulent scale $\eta_{min}$ of the flow scales as $\eta_{min} \varpropto
Re^{-3/4}$. For flows with large enough Reynolds numbers (many astrophysical and geophysical flows reach
Reynolds numbers over $10^{10}$), the smallest turbulent scales have $S_\eta$ smaller than $1$, even
for small grains. This is a well-known behavior of particle-laden flows. Particles are centrifuged out of
the turbulent eddies when their scale Stokes number reaches unity \citep{Squires91}. This creates a converging flow of
particles towards regions of low vorticity. Consequently, this induces a velocity dispersion inside the
particle flow and, for similar reasons to those described in this paper, the particle fluid develops a
pressure. As a side note, when turbulence modelling is considered, turbulent pressure terms can arise from
fluctuating or subgrid velocity dispersions \citep{Dobrovolskis99}. However, turbulent pressure terms are
a consequence of Reynolds averaging or spatial filtering and disappear when turbulence is directly simulated instead of
modelled. \\

This is hopefully a good illustration of the care one has to take whenever a fluid equation is applied to a
collection of solid particles in an astrophysical flow. Whenever any dynamical timescale 
becomes smaller than the characteristic coupling timescale between the particles and the underlying fluid, the
particles cannot be considered as a pressureless fluid, whatever the collision rate between particles.

\begin{acknowledgements}
Part of this work was done during the author's stay at Jeremiah Horrocks Institute for Astrophysics \&
Supercomputing, University of Central Lancashire, 
whose hospitality is gratefully acknowledged. I am grateful to J. Braine, S. Courty, B. Dubrulle, and J.-M.
Hur\'e for comments on the manuscript, and to the anonymous referee and T. Guillot for suggestions that helped me 
to significantly improve the presentation of the paper. 
\end{acknowledgements}

\bibliographystyle{aa}
\bibliography{bib}

\end{document}